\documentclass[floatfix,twoside, twocolumn,aps]{revtex4}

\usepackage{epsfig}

\begin{document}

\title{Modelling Correlations in Portfolio Credit Risk}

\author{Bernd Rosenow\footnote{The first two authors have
contributed equally.}$^1$, Rafael Wei\ss bach$^{\ast 2}$, and
Frank Altrock$^2$}

\address{$^1$ Institut f\"ur Theoretische Physik, Universit\"at zu
  K\"oln, 50923 K\"oln, Germany   \\
  $^2$ Risk Management Support $\&$ Control, WestLB AG, 40217 D\"usseldorf, Germany}

\date{January 19, 2004}

\begin{abstract}
The risk of a credit portfolio depends crucially on correlations between the probability of default (PD) in different economic sectors. Often, PD correlations have to be estimated from relatively short time series of default rates, and the resulting estimation error hinders the detection of a signal. We present statistical evidence that PD correlations are well described by a (one-)factorial model. We suggest a method of parameter estimation which avoids in a controlled way the underestimation of correlation risk. Empirical evidence is presented that, in the framework of the CreditRisk+ model with integrated correlations, this method leads to an increased reliability of the economic capital estimate.

\end{abstract}

\maketitle

Managing portfolio credit risk in a bank requires a sound and
stable estimation of the loss distribution with a special emphasis
on the high quantiles denoted as Credit Value-at-Risk (CreditVaR).
The difference between the CreditVaR and the expected loss has to
be covered by the economic capital, a scarce resource of each
bank.  From a risk management perspective, the definition of
industry sectors allows to diversify credit risk. The degree to
which this diversification is successful depends on the strength
of correlations between the sectors.  Moreover, the correlations
between sector PDs crucially influence the CreditVaR and hence the
economic capital.

In large banks, the concentration risk in industry sectors is a key
risk driver.  Recently, several approaches for describing and
modelling concentration risk were discussed \cite{Bue,NagBa01,Frey+01}.
In CreditRisk+ \cite{CSFB}, concentration risk is modelled as a
multiplicative random effect on the PD per counterpart in a given
sector. In the original version of CreditRisk+, the loss distribution
is calculated for independent sector variables. Correlations between
PD fluctuations in different sectors can be integrated into
CreditRisk+ with the method of B\"urgisser et al. \cite{Bue}.  For the
calculation of the CreditVaR it is important whether input parameters
like the correlation coefficients between sector PDs are known or must
be estimated. In the latter case, this estimation leads to an
additional variability of the target estimate, in our case the
portfolio loss. In this way, uncertainty in the estimation of PD
correlations translates itself into uncertainty of the economic
capital of a bank.

The estimation of cross--correlations is difficult due to the
"curse of dimensionality": if the length $T$ of the available time
series is comparable to the number $K$ of industry sectors, the
number of estimated correlation coefficients is of the same order
as the number of input parameters with the result of large
estimation errors. A way out of this dilemma is the use of a
factor model with a reduced dimensionality of the parameter space.
We present evidence  that the PD correlations for $K=20$ industry
sectors are well captured by a one--factor model. Surprisingly,
even the parameter estimation for the one--factor model is subject
to large statistical fluctuations and gives rise to a considerable
uncertainty in the CreditVaR. We discuss these fluctuations in
detail and suggest a bootstrap method which allows to find an
upper limit for the parameters. We assess the impact of different
conservative estimates with respect to the CreditVaR of a
realistic
portfolio.\\

\noindent
{\bf Description of data set}\\ As the economic activity and the
probability of default in a given industry sector is not directly
observable, we approximate it by the insolvency rate in that sector
over the last $T$ years. The probability of insolvency $PD_{kt}$ of
sector $k$ in year $t$ is calculated as the ratio of the number of
insolvencies in that sector to the total number of companies in the
sector
%
\begin{equation}
\hat{PD}_{k t}= \frac{\sum_{A \rm \; \in \; sector \; k \; in \; year
\; t }I_{ \{A \;\rm  fails\}}}{\sum_{A \rm \; \in \;
sector \; k \; in \; year \; t }   } \ \ .
\label{insolvency.eq}
\end{equation}
%

With the help of insolvency rates, the default probability for a given
company $A$ can be factorized into an individual expected PD $p_A$ and
the sector specific relative PD movement $X_{k}$ with expectation
$\langle X_k \rangle = 1$ according to
%
\begin{equation}
P(A \; \rm fails)=p_A X_k \ \ {\rm with} \ \ X_{k t}=
\frac{\hat{PD}_{k t} }{{1 \over T}\sum_{t} \hat{PD}_{k t}} \ \ .
\end{equation}
%
For this study, we use sector specific default histories as supplied
by the federal statistical office of Germany. We analyze default rates
for a segmentation of the economy into 20 sectors and estimate the
sample covariance matrix $\Sigma^{\rm emp}$ and sample correlation
matrix {\bf \sf C}$^{\rm emp}$ as
%
\begin{equation}
\Sigma_{ij}^{\rm emp} = {1 \over T\! - \!1 } \sum_{t=1}^T (X_{i
t}\! -\! 1) (X_{j t}\! - \!1) , \   C_{ij}^{\; \rm
emp}=\Sigma_{ij}^{ \rm emp} /   \sigma_{X_i} \sigma_{X_j}
\end{equation}
%
with $\sigma_{X_i}^2 = \Sigma^{\rm emp}_{ii}$.\\

\noindent
{\bf Test for independent sectors}\\ We first ask whether the sample
correlation matrix of the  PD time series is compatible with the
hypothesis of zero correlations.  Ideas for testing this hypothesis
for covariance matrices date back to the seventies \cite{JOH}, and
were recently generalized to situations where the number of time
series is larger than the sample size \cite{LED}. Here, we use an
adaption of the tests \cite{JOH,LED} to test for the
equivalence of correlation matrix to the unit matrix. The test
statistics
%
\begin{equation}
R = {1 \over K} {\rm tr}\left[ {\bf \sf C}^2 \right] - 1
 \ ,
\label{rstatistics.eq}
\end{equation}
%
for a correlation matrix {\bf \sf C}
is both $K$-- and $T$--consistent with the $T$--limiting
distribution $(T\!-\!1) K R/2 \stackrel{D}{\rightarrow} \chi^2_{K(K-1)/2}$
\cite{Rosenow03}. The prefactor $T-1$ rather than $T$ is chosen to
improve the finite $T$ properties of the test.  For our example with
$T=7$ and $K = 20$, we find $R=5.805$, whereas the critical value for
$\alpha = 0.05$ is $R_{\rm crit} = 3.719$.  Hence, the independence
of sector PDs must not be assumed and a model describing sector
correlations is needed.\\

\noindent
{\bf Description of one--factor model}\\
We diagonalize the empirical cross correlation matrix {\bf \sf
  C}$^{\rm emp}$ and rank order its eigenvalues $\lambda_{i,{\rm emp}}
< \lambda_{i+1,{\rm emp}}$.  As we are interested in modelling
correlations rather than covariances, we normalize the $X_{it}$
such that they have the same, namely the average variance
$\sigma_X^2 = (1/K) \sum_{i=1}^K \sigma_{X_i}^2$ and subtract the
mean
%
\begin{equation}
\tilde{X}_{it} = (X_{it} -1) {\sigma_X \over \sigma_{X_i}} \ .
\end{equation}
%
We use the components of the eigenvector ${\bf u}^{(K)}_{\; \rm emp}$
corresponding to the largest eigenvalue $\lambda_{K,{\rm emp}}= 10.38$
to define a factor time series
%
\begin{equation}
Y_t = \sum_{i=1}^K u^{(K)}_{i,{\rm emp}}  \tilde{X}_{it}  \ \ .
\label{factorseries.eq}
\end{equation}
%
As compared to averaging the sector variables without prior
information, the definition of the factor time series from the
eigenvector with largest eigenvalue makes sure that the factor
explains a maximum amount of correlations. The idea arises from
factor analysis, see e.g. \cite{Mar}. In the context of stock
returns, a time series defined according to the prescription of
Eq.~\ref{factorseries.eq} was found to agree well with a value
weighted stock index \cite{Ple}. We expect that the factor time
series Eq.~\ref{factorseries.eq} describes economy wide changes of
relative PD, possibly weighted by the economic relevance of the
individual sectors.

We model the correlations between relative PD movements by a
one--factor model
%
\begin{equation}
\tilde{X}_{it}  = b_i Y_t + \epsilon_{it} \ \ .
\label{factor.eq}
\end{equation}
%
The coefficients $\{b_i\}$ are found by performing a linear
regression. To see whether a one--factor model fully describes the
correlations between the $\{\tilde{X}_{it}\}$, we apply the test
Eq.~\ref{rstatistics.eq} to the correlation matrix of the residuals
$\{ \epsilon_{it}\}$. Taking into account that the regression reduces
the effective length of the residual time series from $T$ to $T-1$, we
find $R=4.409$ slightly below the threshold $R_{\rm crit}= 4.463$.
As the assumption of uncorrelated residuals is not rejected, no
further factors are needed for the description of correlations.

\begin{figure}
\centerline{\epsfig{file=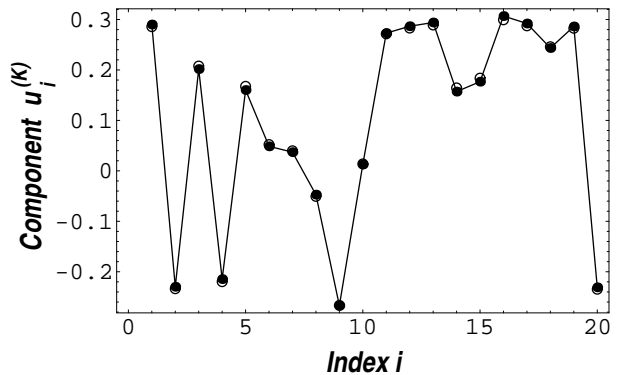,width=8cm}}
\caption{The components of the  eigenvector ${\mathbf u}^{(K)}_{\rm emp}$
of the empirical correlation matrix (connected full circles) are almost
identical to the components of the eigenvector ${\mathbf u}^{(K)}_{\rm
point}$ of the point estimator {\bf \sf C}$^{\rm point}$ (open
circles) .}
\label{pointestimate.fig}
\end{figure}

The point estimator can now be calculated under the assumption
that the residua $\{ \epsilon_{i,t}\}$ are iid observations from
uncorrelated random variables $\epsilon_i$ $i=1, \ldots, K$, i.e.
$\langle \epsilon_i \epsilon_j \rangle = 0$ for $i \neq j$.
Defining the factor variance $\sigma_Y^2 = {1 \over T - 1}
\sum_{t=1}^T Y_t^2$, one finds the point estimator for the cross
correlation matrix as
%
\begin{equation}
C^{\rm point}_{ij}= \delta_{ij} + (1 - \delta_{ij}) b_i b_j
\sigma_Y^2 /  \sigma_{X}^2  \ \ .
\end{equation}
%
The largest eigenvalue of {\bf \sf C}$^{\rm point}$ is found to be
$\lambda_{K,{\rm point}} = 10.66$ in good agreement with the original
largest eigenvalue. In addition, the corresponding eigenvector ${\bf
  u}^{(K)}_{\; \rm point}$ is found to be very close to the
original eigenvector (Fig.\ref{pointestimate.fig}).\\

\noindent
{\bf Fluctuations in empirical correlation matrices -- a toy model}\\

In this section, we use the results of Monte Carlo simulations to study the
relation between the cross correlation matrix {\bf \sf C}$^{\rm
  model}$ resulting from infinitely long model time series and
matrices {\bf \sf C}$^{\rm sim}$ numerically calculated from finite
time series of length $T$. We find that the $\{ {\bf \sf C}^{\rm
  sim}\}$ differ from {\bf \sf C}$^{\rm model}$ both in a systematic
way, for example a shift of the largest eigenvalue towards larger
values, and a random way, i.e.~an individual member of the simulated
ensemble deviates significantly from the average \cite{Laloux+99,Plerou+99}.

To simplify simulations, we rewrite Eq.~\ref{factor.eq} as
%
\begin{equation}
\tilde{X}_{it} = \alpha \beta_i F_t + c_i \eta_{it} \ \ .
\label{simulate.eq}
\end{equation}
%
The random variables are rescaled according
to $c_i \eta_{it} = \epsilon_{it}$ and $\alpha \beta_i F_t = b_i Y_t$
such that their variances are ${\rm var}(F)= \sigma_X^2$ and ${\rm
  cov}(\eta_i \eta_j) = \sigma_X^2 \delta_{ij}$. In addition, the
$\{ \beta_i \}$ obey the normalization condition $\sum_{i=1}^K
\beta_i^2 = 1$, which makes them comparable to the components of
the eigenvector ${\mathbf u}^{(K)}$. The model parameters are
subject to the constraint $c_i^2 = 1 - \alpha^2 \beta_i^2$ in
order to enforce ${\rm var}(\tilde{X}_{it})= \sigma_X^2$.

The model is completely defined by i) the parameter $\alpha$
determining the largest eigenvalue, ii) the parameters $\{ \beta_i\}$
and iii) by the probability distribution of the random variables $F$
and $\{ \eta_i \}$. As the empirical PD movements are commonly assumed
to follow a gamma distribution, we model the random variables $F$ and
$\eta_i$ by gamma distributions as well \cite{gammadistribution}.  In
addition, we use normal distributions for the random variables and
find that the deviation from a simulation with gamma distributed
variables is smaller than $3\%$.  As the simulation of gaussian random
variables is computationally much more efficient than the simulation
of gamma distributed variables, we use normally distributed variables
for the computationally demanding selfconsistent calculations
described in the next section.

\begin{figure}
\centerline{\epsfig{file=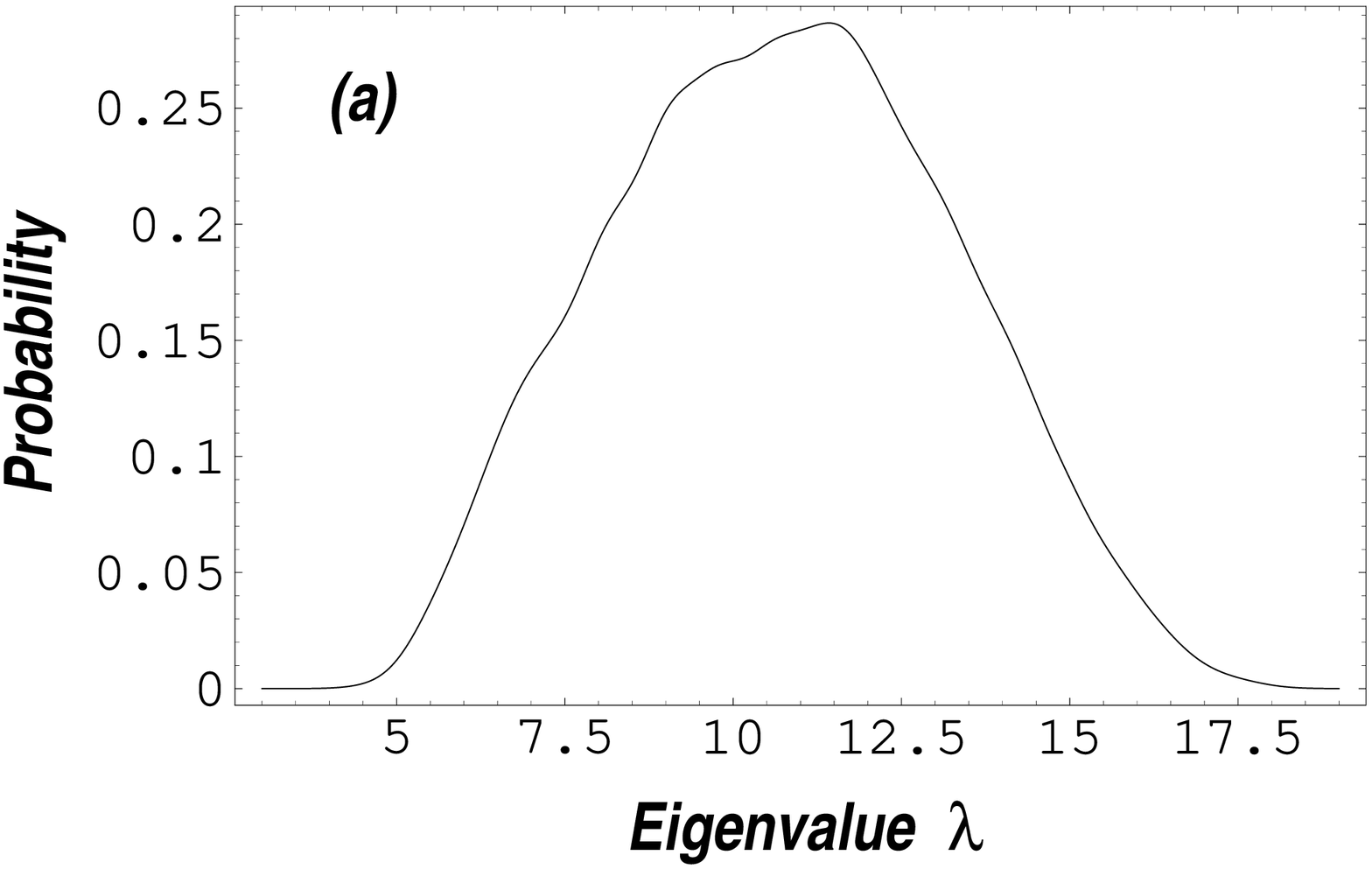,width=8cm}}
\centerline{\epsfig{file=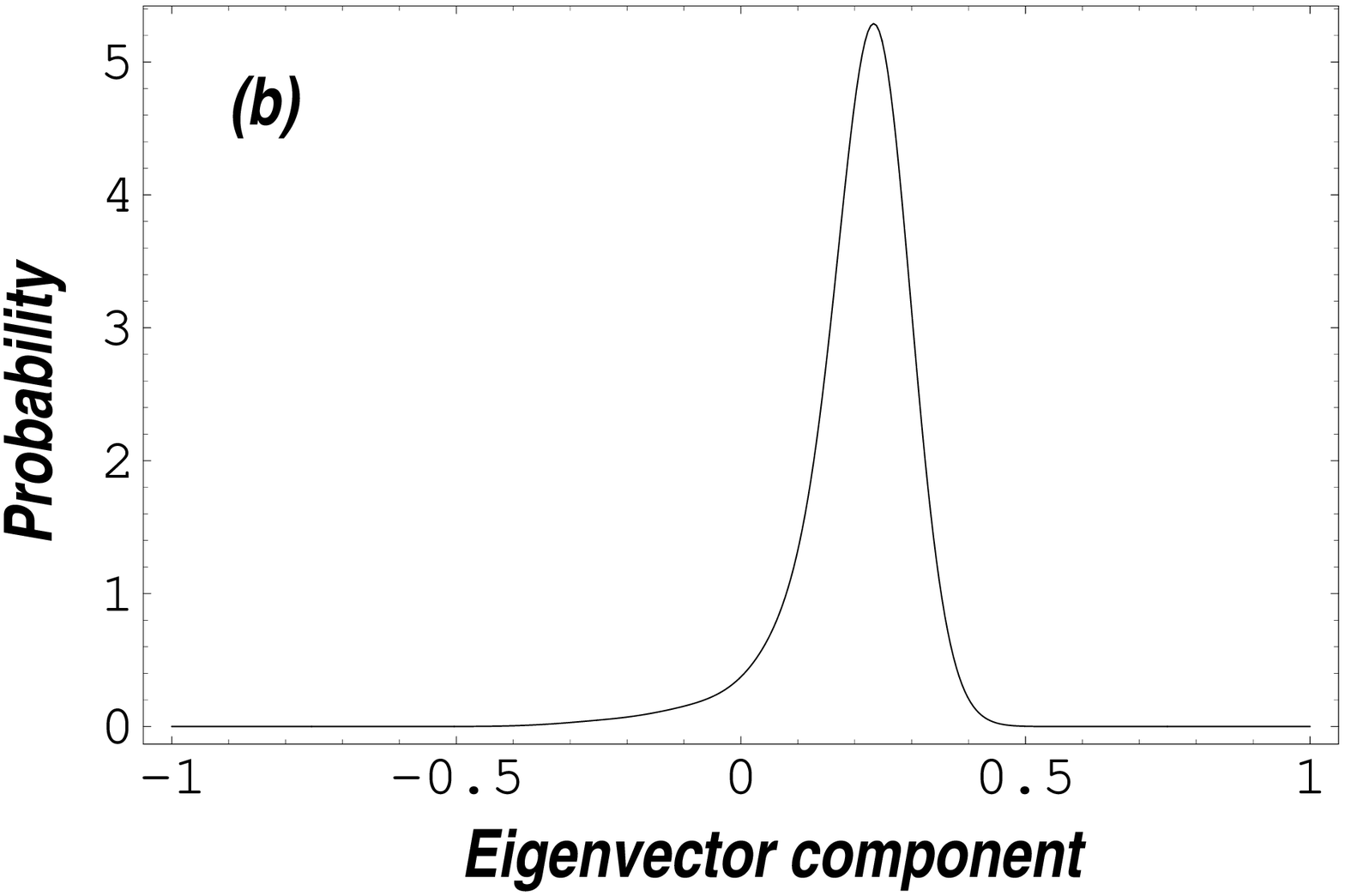,width=8cm}}
\caption{Distribution of (a) the largest eigenvalue and of (b) all
components of the corresponding eigenvector from simulations of
the one--factor toy model with $\lambda_{K,{\rm model}}=10.38$.}
\label{distributions.fig}
\end{figure}

The infinite time series correlation matrix of the model is given by
%
\begin{equation}
C_{ij}^{\rm model} = \delta_{ij} + (1 - \delta_{ij}) \alpha^2 \beta_i
\beta_j \ .
\label{modelpoint.eq}
\end{equation}
%
In this section, we study the outcome of model simulations for the
particularly simple hypothetical case $u^{(K)}_{i, \rm
model}=\beta_i \equiv 1 / \sqrt{K}$ in order to gain qualitative
insight into the occurring fluctuations. For a given value of
$\alpha$ corresponding to $\lambda_{K,{\rm model}}=10.38$, we
perform 10000 Monte Carlo simulations of Eq.~\ref{simulate.eq}.  We
compute the pdf for the largest eigenvalue $\lambda_{K,{\rm sim}}$
of {\bf \sf C}$^{\rm sim}$ and the corresponding eigenvector
${\mathbf u}^{(K)}_{\rm sim}$.  We find that both quantities have
broad distributions (see Fig.~\ref{distributions.fig}). The
distribution of eigenvalues has an average $\langle
\lambda_{K,{\rm sim}} \rangle = 10.72$, which is significantly
larger than the true eigenvalue $\lambda_{K,{\rm model}}=10.38$.
In addition, one finds simulated eigenvalues as low as
$\lambda_{K,{\rm
    sim}}=5$.  We quantify the systematic shift of eigenvalues by the
average $\Delta \lambda = \langle \lambda_{K,{\rm sim}}\rangle -
\lambda_{K,{\rm model}}$.  The magnitude of eigenvalue fluctuations is
described by the standard deviation
%
\begin{equation}
\sigma_\lambda = \sqrt{ \langle \lambda_{K,{\rm sim}}^2\rangle - \langle
\lambda_{K,{\rm sim}}\rangle^2  } \ \ .
\end{equation}
%
For the distribution shown in Fig.~\ref{distributions.fig} we find
$\sigma_\lambda = 2.42$.

There are significant fluctuations of eigenvector components as
well. For theoretical eigenvector components $u^{(K)}_{i \rm
model}= 0.224 \; \forall \; i$ one even finds negative empirical
components indicating spurious anticorrelations, which would lead
to dangerous hedges in credit portfolios. Specifically, we
calculate the standard deviation
%
\begin{equation}
\sigma_{u_i} =\sqrt{ \langle (u^{(K)}_{i,{\rm sim}})^2 \rangle - \langle
u^{(K)}_{i,{\rm sim}}\rangle^2       }
\end{equation}
%
and find $\sigma_{u_i} = 0.083$. Since the $u^{(K)}_{i, \rm
model}$ do not vary across $i$ we only need to estimate one
$\sigma_{u_i}$.

As a conclusion, even if the generating process for relative PD
movements is a simple one--factor model, the empirically found
parameters can deviate significantly from the theoretical ones.
We advocate the point of view that the empirical {\bf \sf C}$^{\rm
emp}$ has to be viewed as a member of such a fluctuating ensemble
in that its eigenvalues and eigenvectors can deviate significantly
from the unknown ``true'' correlation matrix of PD movements
\cite{Laloux+99,Plerou+99}. Then, the statistical properties of
the ensemble $\{ {\bf \sf C}^{\rm sim}\}$ can be used to derive
error bars for both the largest eigenvalue and the components of
the
corresponding eigenvector.\\

\noindent {\bf Conservative estimates}\\ How can we use these
results to make a reliable estimate for the correlation matrix of
relative PD movements? A bank needs to act in a conservative manner to
prevent insolvency. Using the empirical correlation matrix, the bank
risks that the correlations are "accidently" low.  The most
conservative approach would be to assume all correlations to be $1$,
i.e. $u^{(K)}_i=\frac{1}{\sqrt{K}} \; \forall \; i$.  But now the
model would effectively be a one-sector model. Any possibility to
measure concentration risk in certain industry sectors would be
prevented. The model would not encourage diversifying the business
across sectors.

As a controlled mediation we introduce "cases" of add-ons of
$x=1,2,3$ standard deviations to the fluctuating quantities such
that the predicted risk for a portfolio is increased. This means
correcting the eigenvalue towards larger values and the
eigenvector components towards the value $u^{(K)}_i \equiv
1/\sqrt{K}$ indicating the same correlation strength for all
sectors and the absence of hedge possibilities.

\begin{figure}
\centerline{\epsfig{file=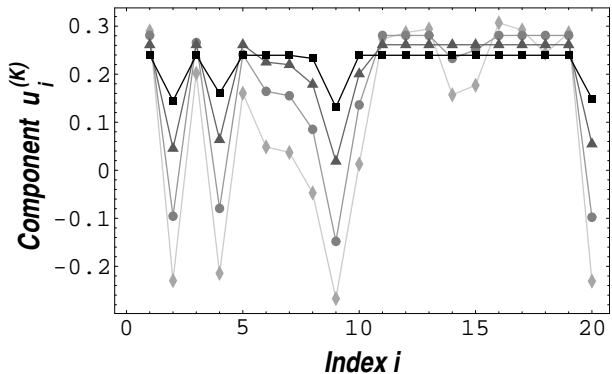,width=8cm}}
\caption{Comparison between the empirical eigenvector ${\mathbf
u}^{(K)}_{\rm emp}$ (diamonds) and the conservative estimates
${\mathbf u}^{(K)}_{1 \sigma}$ (circles), ${\mathbf u}^{(K)}_{2
\sigma}$ (triangles), and ${\mathbf u}^{(K)}_{3 \sigma}$
(squares).} \label{conservative.fig}
\end{figure}

Specifically, we let $u^{(K)}_{i,{\rm case}}= 1/\sqrt{K}$ if
%
\begin{equation}
|u^{(K)}_{i,{\rm emp}} - 1/\sqrt{K}| < x \cdot \sigma_{u_i}
\label{vectorequation.eq}
\end{equation}
%
and $u^{(K)}_{i,{\rm case}}= u^{(K)}_{i,{\rm emp}} \pm x \cdot
\sigma_{u_i}$ otherwise. The sign is chosen such that the overall
risk increases, i.e. such that $u^{(K)}_{i,{\rm case}}$ falls
between the empirical value and $1/\sqrt{K}$. After applying these
corrections, the eigenvector is normalized. For this calculation,
we fix the parameter $\alpha$ such that the simulated largest
eigenvalue $\lambda_{K,\rm sim}$ is equal to the empirically
observed one. We calculate the $\{\sigma_{u_i}\}$
selfconsistently, i.e.~we calculate $\sigma_{u_i}$ for the
$u^{(K)}_{i,\rm case}$ which solves Eq.\ref{vectorequation.eq}.
The results are shown in Fig.\ref{conservative.fig}. We see that
for increasing $x=1, 2, 3$, the model eigenvector comes closer to
the null hypothesis of an eigenvector with identical components.
While the empirical eigenvector has significant negative
components indicating anticorrelations between some of the
sectors, the negative components in ${\mathbf
  u}^{(K)}_{1 \sigma}$ are already strongly reduced and completely
gone in ${\mathbf u}^{(K)}_{2 \sigma}$.

Similarly, we add a fluctuation margin to the model eigenvalue
\cite{eigenvalueshift} such that
%
\begin{equation}
\lambda_{K,{\rm case}} =  \lambda_{K,{\rm emp}} + x \cdot
\sigma_\lambda    \ \ . \label{lambdaequation.eq}
\end{equation}
%
Here, $x$ specifies the width of the confidence interval for the
estimation of $\lambda_{K,{\rm model}}$. We perform this
calculation selfconsistently, i.e.~we calculate $\Delta \lambda$
and $\sigma_\lambda$ for the $\lambda_{K,{\rm case}}$ which solves
Eq.\ref{lambdaequation.eq}. We find $\lambda_{K, 1\sigma}=11.17$,
$\lambda_{K, 2\sigma}=12.93$, and $\lambda_{K, 3\sigma}=15.42$.\\

\noindent
{\bf Economic implications of the different correlation matrices}\\ So
far, we have described five different estimates for the cross
correlation matrix, i.e.  {\bf \sf C}$^{\rm emp}$, {\bf \sf C}$^{\rm
point}$, {\bf
\sf C}$^{\rm model}_{1 \sigma}$, {\bf \sf C}$^{\rm model}_{2 \sigma}$,
and {\bf \sf C}$^{\rm model}_{3 \sigma}$. To judge the economic
implications of these estimates, we study the differences in the loss
distribution resulting from these correlation estimations. The
CreditVaR is a key quantity in banking when it comes to risk
management. Reduced by the expected loss, it quantifies the capital
needed to prevent insolvency for a given level of security. As capital
is a resource it must be considered in the pricing of credit and
trading products. Therefor, we quantify the impact of the different
correlation estimates by calculating their influence on CreditVaR.

\begin{table}[t]
\begin{tabular}{|c|c|} \hline
correlation matrix & CreditVaR (in billion Euro)     \\ \hline
{\bf \sf C}    &    2.872\\ \hline
{\bf \sf C}$^{\rm point} $   &  2.825 \\ \hline
{\bf \sf C}$^{\rm model}_{1 \sigma}$   & 3.172\\ \hline
{\bf \sf C}$^{\rm model}_{2 \sigma}$  & 3.465   \\ \hline
{\bf \sf C}$^{\rm model}_{3 \sigma}$  & 3.665    \\ \hline
\end{tabular}
\caption{ Analysis of CreditVaR for different correlation matrices    }
\label{table1.tab}
\end{table}

The portfolio we study is realistic -- although fictitious -- for an
international bank. It consists of 4934 risk units distributed
asymmetrically over 20 sectors with 20 to 500 counterparts per sector.
The total exposure is 70 bn Euro with a largest exposure of 1.5 bn
Euro and a smallest exposure of 0.25 mn Euro.  The counterpart
specific default probability varies between 0.03\% and 7\%, the
expected loss for the total portfolio is 373.3 mn Euro.  Table I shows
the CreditVaR calculated by using CreditRisk+ and the method of
B\"urgisser et al. \cite{Bue} for integrating correlations.

We note that the use of a one--factor model changes the CreditVaR only
by two percent as compared to the sample cross correlation matrix.
Thus, the assumption of a one--factor description and the increase of
estimation confidence achieved with this assumption yields portfolio
risk estimation compatible with the parameter free estimation.

Our aim is to estimate a quantile of a probability distribution --
namely the CreditVaR of the portfolio loss distribution. In the
presence of an unknown parameter, it is a well established
statistical result (see \cite{LEH}) that the use of the point
estimate for the parameter -- derived by a model or not -- leads
to an underestimation of the quantile estimate. To account for
this additional estimation insecurity, we add a volatility
$\sigma$ to the parameter estimate, i.e. the correlation matrix.
When applying a one--$\sigma$ estimate, the CreditVaR increases by
400 mn Euro, for the two--$\sigma$ estimate there is another
increase by 300 mn Euro, and using the three--$\sigma$ estimate
the CreditVaR increases by yet another 200 mn Euro. To put these
numbers in perspective, we note that the CreditVaR without
including correlations is found to be 2.27 bn Euro, and that the
assumption of full correlations among all sectors leads to a
CreditVaR of 3.952 bn Euro. Because negative PD correlations are
not plausible from an economic point of view, the use of the
two--$\sigma$ estimate guarantees a sufficient forecast
reliability on the one hand and allows for some guidance for
economical decision on the other hand.

In summary, we have shown that correlations between empirical default
rates for economic sectors are statistically significant and must be
taken into account. We have described these correlations with a
one--factor model and found that this description reproduces well the
empirical correlations. However, when using the model to generate
short time series and calculating their correlation matrix, one
typically observes large statistical fluctuations in the correlation
structure. Due to these fluctuations, the parameter estimation for a
one--factor model is plagued by large uncertainties. When estimating
the model parameters in such a way that the empirically observed ones
appear as a worst case scenario, the reliability of the estimate is
increased in a systematic way, leading to a moderately increased
CreditVaR.

{\it Acknowledgement:} We would like to thank A.~M\"uller-Groeling for
initiating this project. We thank  S.~L\"osch, C.~von Lieres, 
and A.~Wilch for useful discussions.


\begin{thebibliography}{99}
\bibitem{gammadistribution} As gamma distributed variables have nonzero
mean, the model must be rewritten as $\tilde{X}_{it} + 1 = \alpha
\beta_i F_t + c_i \eta_{it}$, i.e.~variables $\tilde{X}_{it} + 1$ with
the original expectation value one are modelled. The expectation
values of the random variables are $\langle F \rangle = 1/( \alpha {1
\over K} \sum_{i=1}^K \beta_i + {1 \over K} \sum_{i=1}^K c_i)$ and
$\langle \eta_i \rangle = (1 - \alpha \beta_i \langle F \rangle )/c_i$
such that $\langle \tilde{X}_i \rangle = 0$ for all $i$.
\bibitem{eigenvalueshift}  For the numerical calculation, we also
account for the systematic eigenvalue shift $\Delta \lambda$.
\end{thebibliography}
\end{document}